\documentclass[letterpaper,superscriptaddress,english,aps,prl,twocolumn,floatfix, showpacs, amsfonts, amssymb]{revtex4}
\usepackage{epsfig, graphicx,psfrag,amsmath,amssymb,float}
\usepackage[T1]{fontenc}
\usepackage[latin9]{inputenc}
\usepackage{amsmath}
\usepackage{amssymb}

\usepackage{amscd}
\usepackage{bm}
\usepackage{psfrag}

\usepackage{babel}

\newcommand{\be}{\begin{equation}}
\newcommand{\ee}{\end{equation}}
\newcommand{\bea}{\begin{eqnarray}}
\newcommand{\eea}{\end{eqnarray}}

\begin{document}
\title{Exact crossover Green function in the two-channel
and two-impurity Kondo models}

\author{Eran Sela, Andrew K. Mitchell and Lars Fritz}

\affiliation{Institute for Theoretical Physics, University of Cologne, 50937 Cologne, Germany}

\begin{abstract}
Symmetry-breaking perturbations destabilize the critical points of the
two-channel and two-impurity Kondo models,
thereby leading to a crossover from non-Fermi liquid behavior to
standard Fermi liquid physics. Here we use an analogy between
this crossover and one occurring in the boundary Ising model to
calculate the full crossover Green function analytically. In remarkable
agreement with our numerical renormalization group calculations, the
single exact function applies for an arbitrary mixture of the relevant
perturbations in each model. This rich behavior resulting from finite
channel asymmetry, inter-lead charge transfer and/or magnetic field
should be observable in quantum dot or tunneling experiments.
\end{abstract}

\maketitle

The most basic quantum impurity model exhibiting non-Fermi
liquid (NFL) behavior is arguably the two-channel Kondo (2CK)
model~\cite{NozieresBlandin}, describing the symmetric
antiferromagnetic coupling of a local spin-$\tfrac{1}{2}$ impurity to
two equivalent but independent conduction channels. The resulting
ground state possesses various intriguing properties, including
notably a residual entropy~\cite{s} of $\tfrac{1}{2}k_B\ln(2)$ and
conductance that approaches its
$T=0$ value as
$\sqrt{T}$ (for a review see Ref.~[\onlinecite{rev:cox_zaw}]).

The same behaviour is predicted at the critical point of the
two-impurity Kondo (2IK) model~\cite{CFT2IKM}. The
tendency to form a trivial local singlet state is favoured by an
exchange coupling acting directly between the impurities; while
the coupling of each impurity to its own metallic lead favours
separate single-channel Kondo screening. The resulting competition
gives rise to a critical point that is closely related to
the 2CK state.

The central difficulty in realizing experimentally the NFL physics of
either model is the extreme delicacy of the 2CK fixed point (FP) to
various symmetry-breaking perturbations. Channel asymmetry, magnetic
field and inter-lead charge transfer processes all destabilise the 2CK
FP and destroy NFL behaviour in both 2CK and 2IK models. 

Tremendous efforts have been made to suppress these relevant
perturbations in order to observe the characteristic NFL
behaviour in a real 2CK device. The quantum dot system realized recently in
Ref.~\cite{Potok07} shows unambiguous signatures associated with
flow to the 2CK FP. But in any real system, the presence of
destabilising perturbations is totally inevitable, leading ultimately
to a crossover from NFL behaviour to standard Fermi liquid physics on
the lowest energy scales.

In this Letter we demonstrate that an arbitrary mixture of the relevant
perturbations in either the 2CK or 2IK model leads to
low-energy behaviour of the impurity Green function that is wholly
characteristic of the incipient 2CK state, and which can not be extracted from the Bethe-ansatz solution~\cite{andrei}. We derive a
single exact function to describe this crossover, which agrees
perfectly with our full numerical renormalization group (NRG) calculations for
both models, and whose rich behaviour should be directly observable in
quantum dot or tunneling experiments.

\textit{Models and perturbations.--}
We consider the standard 2CK and 2IK models,
\begin{align}
\label{2ck}
H_{2CK} =H_0+&J  \vec{S} \cdot (\vec{s}_{0L} + \vec{s}_{0R})  +\delta
H_{2CK},\\
H_{2IK} =H_0+&J(  \vec{S}_L \cdot \vec{s}_{0L} + \vec{S}_R
\cdot \vec{s}_{0R} ) +K \vec{S}_L \cdot \vec{S}_R + \delta H_{2IK}, \nonumber \\
\end{align}
where $H_{0}=\sum_{\alpha,k} \epsilon_{k}^{\phantom{\dagger}} \psi_{k}^{\dagger
  \sigma \alpha} \psi^{\phantom{\dagger}}_{k \sigma\alpha}$ describes two free
conduction electron channels $\alpha=L/R$, with spin density
$\vec{s}_{0\alpha}=\sum_{\sigma\sigma'}\psi_{0}^{\dagger
  \sigma \alpha}
(\tfrac{1}{2}\vec{\sigma}_{\sigma\sigma'})\psi^{\phantom{\dagger}}_{0
  \sigma' \alpha}$ (and $\psi_{0}^{\dagger \sigma \alpha}= \sum_{k}\psi_{k}^{\dagger
  \sigma \alpha}$) coupled to one spin-$\tfrac{1}{2}$ impurity
$\vec{S}$ (2CK) or two impurity spins $\vec{S}_{L,R}$ (2IK).
For $\delta H_{2CK}=0$, the ground state
of $H_{2CK}$ is described by the 2CK FP. Likewise, a critical
inter-impurity coupling $K_c$ can be found such that the ground state
of $H_{2IK}$ is similarly described by the 2CK FP for $\delta H_{2IK}=0$.

Relevant perturbations to each model have been identified from
conformal field theory (CFT)~\cite{CFT2CK,CFT2IKM} and are generically
present in experiment. Specifically,

\be
\label{delta2CK}
\delta H_{2CK} =\sum_{\ell = x,y,z} \Delta_\ell \sum_{\alpha,\beta}\sum_{\sigma\sigma'}
\psi_{0}^{\dagger \sigma \alpha} (\tfrac{1}{2}\vec{\sigma}_{\sigma\sigma'}
\tau_{\alpha\beta}^{\ell})\psi_{0 \sigma'\beta}  \cdot \vec{S} +
\vec{B} \cdot \vec{S},
\ee
describes $L/R$ channel asymmetry in the 2CK model for $\Delta_z\ne
0$, while charge transfer between the leads is embodied in the
$\Delta_x$ and $\Delta_y$ components of the first term [here
$\vec{\tau} ~(\vec{\sigma})$ are the Pauli matrices in the channel (spin) sector]. The second
term describes a magnetic field acting on the impurity.
For the 2IK model, the critical point is destabilized by finite
$(K_c-K)$, and also through
\be
\label{delta2IK}
\delta H_{2IK} = \sum_{\sigma}(V_{LR} \psi_{0}^{\dagger \sigma L} \psi_{0 \sigma R} +
\text{H.c.}) + \vec{B}_s \cdot (\vec{S}_L - \vec{S}_R),
\ee
where $V_{LR}$ describes electron tunneling between the leads
and $\vec{B}_s$ the application of a staggered magnetic field.

\begin{figure*}[t]
\begin{center}
\includegraphics*[width=170mm]{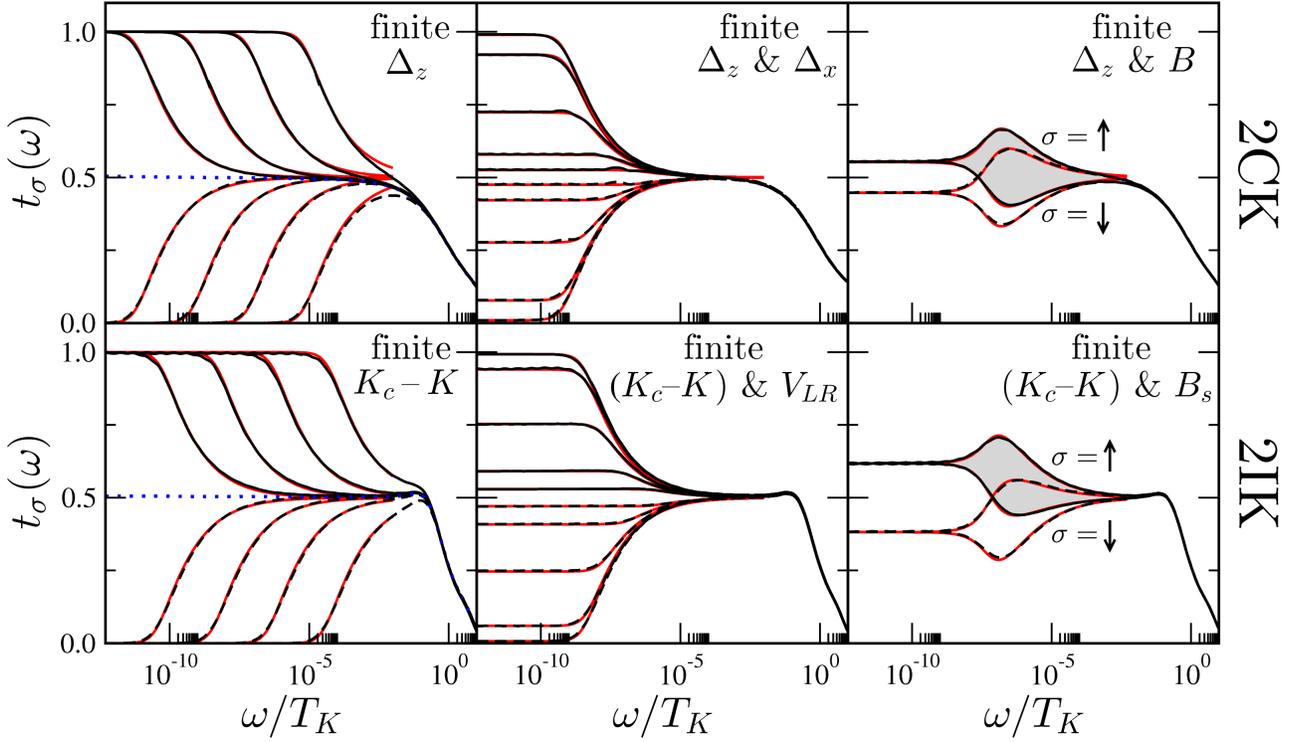}
\caption{\label{fig}
Spectrum $t_{\sigma}(\omega)$ vs. $\omega /T_K$ for the 2CK
model (upper panels) and the 2IK model (lower panels) at $T=0$ in the
presence of various perturbations. Entire frequency dependence
calculated by NRG (black lines); low-energy $\omega \ll T_K$
behaviour in each case compared with exact crossover function
Eq.~\eqref{exactresult} (red lines). All results presented for $\nu J=0.25$.
\emph{Left}: effect of channel asymmetry
$\Delta_z \ne 0$ (2CK) or deviation from critical coupling $(K_c-K)
\ne 0$ (2IK). Specifically, $4\nu\Delta_z = (K_c-K)/D= \pm
10^{-n}$ (with $\pm$ for solid and dashed lines, respectively), and
$n=3,\tfrac{7}{2},4,\tfrac{9}{2},5$ in order of decreasing $T^{*}$,
approaching successively the limit $\Delta_z=(K_c-K)=0$ (dotted line).
\emph{Centre}: effect of including also finite left/right tunneling,
$\Delta_x\ne 0$ (2CK) or $V_{LR}\ne 0 $ (2IK). Shown for fixed finite
$4\nu\Delta_z = (K_c-K)/D=\pm 10^{-5}$ with $5\Delta_x/|\Delta_z| =
2D\nu V_{LR}/|K_c-K| = 10^{-m}$ with $m=2, \tfrac{3}{2}, 1, \tfrac{1}{2},
0$, successively approaching $t_{\sigma}(0)=\tfrac{1}{2}$ from above
[solid lines; $\Delta_z,(K_c-K)>0$] and from below [dashed lines;
$\Delta_z,(K_c-K)<0$].
\emph{Right}: effect of including finite magnetic field. Shown again
for $4\nu\Delta_z = (K_c-K)/D=\pm 10^{-5}$, but now with $B/4D\nu |\Delta_z| =
B_s/|K_c-K| = 10^{1/2}$. As before, solid lines for $\Delta_z,(K_c-K)>0$ and
dashed lines for $\Delta_z,(K_c-K)<0$, with both $\sigma=\uparrow$ and
$\downarrow$ spectra shown. Excellent agreement between NRG data and analytic
curves obtained in all cases from a single set of fitting parameters
for each model: $c_T=96$ and $c_B=0.04$ for 2CK;
while $c_T=0.63$, $c_V=2.4$, $c_B=1.3$ for 2IK.
}
\end{center}
\end{figure*}

These perturbations generate a new energy
scale, $T^{*}$, characterizing the flow away from the 2CK FP, and toward
the Fermi liquid (FL) FP. Signatures of this crossover appear
in the energy-resolved local density of states, since inelastic
scattering ceases at energies $\ll T^{*}$ where the impurity
degrees of freedom are quenched. Indeed, the $dI/dV$
conductance through a quantum dot
asymmetrically coupled to source and drain leads
at zero temperature is related~\cite{asymmetric} to the scattering
$T$-matrix:
$dI/dV\propto \sum_{\sigma=\uparrow,\downarrow} [- \pi \nu
{\rm{Im}} T_{L \sigma}(eV)]$, where $\nu$ is the
lead density of states per spin  and $V$ the source-drain voltage. The Green function is given~by
\bea
\label{tmatrixdef}
G^{\alpha\sigma}_{k k'}(\omega) = G_{k}^0(\omega) \delta_{k k'} +G_{k}^0(\omega) T_{\alpha\sigma}(\omega) G_{k'}^0(\omega),
\eea
with $G_{k}^0(\omega)=(\omega - \epsilon_k + i 0^{+})^{-1}$. Conductance measurements of the 2CK
device of Ref.~\cite{Potok07} or in the proposed 2IK setup of
Ref.~\cite{Zarand2006} thus yield access to the $T$-matrix and
hence the full Green function.
Our goal here is to calculate the quantity $t_{\sigma}(\omega)=- \pi \nu
{\rm{Im}} T_{L \sigma}(\omega)$ exactly and numerically for
the 2CK and 2IK models in the presence of symmetry-breaking
perturbations described by~Eqs.~\eqref{2ck}--\eqref{delta2IK}.

\textit{Results.--}
Preempting the technical discussion of the next section, we present
now the key results of the Letter.
Heralding renormalization group flow to the stable FL FP, the
low-energy crossover scale is given generically by
\be
\label{tstar}
T^{*}=c_T\lambda^2,
\ee
where $\lambda^2= \sum_{j=1}^8 \lambda_j^2$
(and $\{\lambda_4,\lambda_5,\lambda_6
\}=\{\lambda_B^x,\lambda_B^y,\lambda_B^z \}$).
For the 2CK model,
$\lambda_1= \nu\Delta_z\sqrt{T_K}$,
$\lambda_{2,3}= c_V \nu\Delta_{x,y}\sqrt{T_K}$ and
$\vec{\lambda}_B= c_B \vec{B}/\sqrt{T_K}$; while for 2IK,
$\lambda_1= (K_c-K)/\sqrt{T_K}$,
$\lambda_2+i\lambda_3= c_V \sqrt{T_K} \nu V_{LR}$, and
$\vec{\lambda}_B= c_B \vec{B}_s/\sqrt{T_K}$.
Here $c_T, c_V, c_B = \mathcal{O}(1)$ are fitting parameters which
depend on the model and on $J$, and $T_K \propto e^{-\frac{1}{\nu J}}$ is the Kondo temperatre. The perturbations associated with
coupling constants $\lambda_7$ and $\lambda_8$  do not conserve total charge~\cite{CFT2CK,CFT2IKM}, so are ignored.

In the simplest case of channel anisotropy in the 2CK model, the
result $T^{*}\propto (\Delta_z)^2$ has long been
established~\cite{rev:cox_zaw}. The extension to finite
$\lambda_1,\lambda_2,\lambda_3\ne 0$ follows~by trivial rotation of the
bare Hamiltonian in $\vec{\tau}$-space, implying directly that $T^{*}\propto
(\Delta_x)^2+(\Delta_y)^2+(\Delta_z)^2$, and~hence $c_V=1$ for the 2CK
model. However, the
low-energy effective Hamiltonian for both 2CK and 2IK models possesses
a larger SO$(8)$ \emph{emergent} symmetry
that
permits a similar rotation, yielding the
generalization, Eq.~\eqref{tstar}. Importantly, we show that the same
rotation can also be exploited to obtain a single zero temperature Green function for
the crossover. Our exact result for the $T$-matrix is
\begin{equation}
\label{exactresult}
2 \pi i \nu T_{\alpha\sigma}(\omega) =   1  -S_{\alpha\sigma} \mathcal{G} \left( \omega/T^* \right),
\end{equation}
where
$\mathcal{G}(x) = \frac{2}{\pi} K\left[i x  \right]$, $K[x]$ is the complete elliptic integral of the first
kind, yielding asymptotically $\mathcal{G}[x] = 1+ i x/4-(3
x/8)^2+\mathcal{O}(x^3)$ for $x \ll 1$; and $\mathcal{G}[x] =
\frac{\sqrt{i}}{2\pi}(\log[256 x^2]-i  \pi)x^{-1/2}$
for $x \gg 1$. $S_{\alpha\sigma}$ is the FL S-matrix, containing phase shift information, given by
\be
\label{Smatrix}
S^{2CK}_{\alpha\sigma} = (-\alpha\lambda_1+i\sigma
  \lambda_B^z)/\lambda= \alpha S^{2IK}_{\alpha\sigma},
\ee
with $\sigma=\pm 1$ for spins $\uparrow/\downarrow$ and $\alpha=\pm 1$
for channel $L/R$, such that $t_\sigma(0)=\frac{1}{2}-\frac{1}{2}{\rm{Re}} S_{L \sigma}$ (and we have used $\vec{\lambda}_B \parallel \hat{z}$).

Rich physical behaviour is obtained when combinations of the relevant
perturbations are applied. We now discuss relevant and representative
cases of the general and exact formula, Eq.~\eqref{exactresult}, valid
in the crossover regime $|\omega|\ll T_K$. We also
employ the NRG technique (for a review, see
Ref.~\cite{nrg:rev}), which can be used to determine
accurately $t_{\sigma}(\omega)$ on all energy scales, for both 2CK and
2IK models in the presence of any perturbation. Here we focus on
the low-energy behaviour where comparison can be made to
the exact results. All symmetries of the
problem are exploited, and $N_s=6000$ states are retained at each
iteration. The leads of width $2D$ are assumed to have a uniform
density of states $\nu=1/(2D)$, and are discretized logarithmically
\cite{nrg:rev}.

NRG data and exact results are presented in Fig.~\ref{fig}, with parameters
given in the caption. $t_{\sigma}(\omega)$ vs. $\omega/T_K$ is
plotted for the 2CK model (upper panels) and the 2IK model (lower
panels) in the presence of various perturbations.
In the left panels, finite $\lambda_1$ is considered (channel
asymmetry for the 2CK model, and detuning of the inter-impurity
coupling in the 2IK model). Precisely at the 2CK FP, \textbf{$\lambda=0$},
$S_{\alpha\sigma}=0$~\cite{CFT2CK,CFT2IKM}, and hence
$t_{\sigma}(\omega)=\tfrac{1}{2}$ for $|\omega|\ll T_K$. But for
$\lambda_1\ne 0$  one immediately obtains
$|t_{\sigma}(\omega) - \tfrac{1}{2}|\sim |\omega/T^{*}|^{-1/2}$
for $T^{*}\ll |\omega|\ll T_K$, while for $|\omega|\ll T^{*}$, the
classic quadratic approach to the FL FP is given asymptotically by
$|t_{\sigma}(\omega)-t_{\sigma}(0)|\sim (\omega/T^{*})^2$, with
$t_{\sigma}=1$ (solid lines: $\lambda_1>0$) or
$t_{\sigma}=0$ (dashed lines: $\lambda_1<0$) being obtained at $\omega=0$.
This full lineshape was likewise obtained numerically for
example in Ref.~\onlinecite{Zarand2006} for the 2IK model, or for the 2CK
model~\cite{Toth,borda} and related odd-impurity chains in
Ref.~\cite{akm:oddimp}.  The exact crossover
function, Eq.~\eqref{exactresult}, is plotted as the red line in each
case, showing remarkable agreement for $|\omega|\ll T_K$.

The effect of including also left/right charge transfer terms (finite
$\lambda_2$) is shown in the centre panels ($\lambda_1$ now being kept
fixed). As $\lambda_2$ increases, the $\omega \ll T_K$ lineshapes seen in the left panels of
Fig.~\ref{fig} undergo a simple rescaling $t_\sigma(\omega) \to \frac{1}{2} + \left| \frac{\lambda_1}{\lambda} \right| [t_\sigma(\omega) - \frac{1}{2}]$ and eventually for $|\lambda_2|\gg
|\lambda_1|$ they completely flatten.
In the 2CK model, the resulting form of $t_{\sigma}(\omega)$
is readily understood: rotation in $\vec{\tau}$-space allows the
Hamiltonian to be written in terms of $\Delta_z$ only --- but
$t_{\sigma}(\omega)$ is then a weighted combination of the
$\Delta_z>0$ and $\Delta_z<0$ spectra shown in the left panels. That
the same behaviour is observed in the 2IK model (lower-centre panel)
is a deeply non-trivial result however. There is no symmetry of the
bare Hamiltonian that permits this rotation; rather, it is the result
of an emergent symmetry. In both cases, the crossover function is
described perfectly by Eq.~\eqref{exactresult}.

Finally, we consider the effect of applying a magnetic field (finite
$\lambda_B^z$). In the absence of other perturbations,
$\text{Re}S_{\alpha\sigma} = 0$, and hence
$t_{\uparrow}(0)=t_{\downarrow}(0)=\tfrac{1}{2}$ (consistent with a $\pi/4$ phase shift~\cite{pustilnik04}).
Indeed, $t_{\sigma}(\omega)=\tfrac{1}{2}$ for
$T^{*}\ll |\omega|\ll T_K$ since the system is near the 2CK
FP. However, the impurity magnetization $M\sim
B^z$ for the 2CK model (or staggered magnetization $M_s\sim B^z_s$ in
the 2IK model); thus $t_{\uparrow}(\omega)\ne t_{\downarrow}(\omega)$
since $M\propto \int_{-\infty}^{0}\text{d}\omega [
t_{\uparrow}(\omega) - t_{\downarrow}(\omega) ]\ne 0$. For finite
(staggered) magnetization therefore, the `up' and `down' spectra must
deviate at finite frequency:
a `pocket' opens between the curves at $|\omega|\sim T^{*}$, whose area
is proportional to the (staggered) magnetization. This behavior is
observed in the right panels of Fig.~\ref{fig}, where a fixed
$\lambda_1$ is also included.

For $|\omega|\ll T^{*}$, we find a \emph{linear} approach to the FL FP,
$t_{\sigma}(\omega)-t_{\sigma}(0)=\frac{\sigma \lambda_B^z  }{8 \lambda} \frac{\omega}{T^*} + \mathcal{O} \left( \frac{\omega^2}{{T^*}^2} \right)$, rather
than the quadratic dependence usually associated with FL theory.
However, this result is a perfectly natural
consequence of the complex scattering matrix
$S_{\alpha\sigma}$. $t_{\sigma}(\omega)$ comprises
contributions from both imaginary and real
parts of the complex function $\mathcal{G}(\omega/T^{*})$ in
Eq.~\eqref{exactresult}, the latter of which contains a leading linear term.
It is this contribution that
of course dominates the low-$|\omega|$ behaviour of
$t_{\sigma}(\omega)$ in the presence of the magnetic field. Again, there
is excellent agreement between the NRG data and the exact results across
the entire frequency range $|\omega|\ll T_K$. We finally note that the
conductance is obtained by averaging $\sigma=\uparrow$ and
$\downarrow$ contributions; the imaginary part of
$S_{\alpha\sigma}$ thus cancels by Eq.~\eqref{Smatrix}, and
the characteristic `hump' observed in $t_{\sigma}(\omega)$ is absent
in $dI/dV$. In consequence, $dI/dV\sim (eV)^2$ for $eV\ll T^{*}$
at finite fields. We now sketch the derivation of Eqs.~\eqref{exactresult}~and~\eqref{Smatrix}.

\textit{Analytic crossover function at $K \ne K_c$ in the 2IK model.--}
The detailed CFT analysis of the 2IK model in
Ref.~\cite{CFT2IKM} demonstrated that perturbing the critical
point by finite $(K_c-K)$ is equivalent to the action of a boundary
magnetic field $h$ in the boundary Ising model (BIM) in 2 dimensions,
described in the field theory limit by~\cite{ghosal}
\be
\label{fieldtheorylimit}
\mathcal{H}_{\textit{Ising}} =\frac{1}{2} \int_{- \infty}^\infty dx \epsilon(x) i \partial_x \epsilon(x) + i h  \epsilon (x=0) a\;.
\ee
$\epsilon(x)$ and $a$ are Majorana fermion (MF) fields ($a$ is local) We use the unfolded coordinate system with left moving
convention, where $x>0$ correspond to incoming fields, $x<0$ to
outgoing fields, and $x=0$ is the boundary itself (for more details
see also Ref.~\cite{Sela2009PRB}). The
renormalization group flow from free to fixed boundary
condition~\cite{cardy} is identical in both 2IK and BIM. The energy
scale associated with the crossover in the BIM is $T^* \propto h^2$.
We now apply this analogy and the machinery of CFT to obtain an exact
result for the crossover Green function in
the 2IK model with finite $(K_c-K)$. The Fourier transform of the
electron Green function in Eq.~\eqref{tmatrixdef} factorizes~into
\bea
\label{g}
\langle \psi_{\sigma \alpha}(z_1) \psi_{\sigma' \beta}^\dagger(\bar{z}_2) \rangle \propto  \frac{\delta_{\sigma \sigma'}  \delta_{\alpha \beta}}{(z_1-\bar{z}_2)^{\frac{7}{8}}} \langle \sigma(z_1) \sigma(\bar{z}_2) \rangle_h,
\eea
where $z_1=\tau+i x_1$ and $\bar{z}_2=-i x_2$. Here $\tau$ is
imaginary time, and $x_1,x_2>0$, implying that the Green function
probes an electron propagating through the boundary. The electron
Green function is thus given in terms of  the two-point function of the
chiral Ising spin-operator $\sigma$, calculated with
Eq.~\eqref{fieldtheorylimit}. The magnetization in the Ising model
evaluated at distance $y$ from the boundary is given
by~\cite{affleckreview} $m(y) = \langle \sigma(z_1) \sigma(z_1^*)
\rangle_h$, where $z_1=i y$.
The full function $m(y)$ at finite $h$ has been calculated exactly by Chaterjee and Zamolodchikov~\cite{cz}:
\be
\label{M}
m(y) \propto (4 \pi h^2 y)^{\frac{3}{8}}e^{4 \pi h^2 y} K_0(4 \pi h^2 y),
\ee
where $K_0$ is the modified Bessel function of the second
kind. Eqs.~\eqref{g}, \eqref{M} thus allow the Green function to be
calculated when $\bar{z}_2 = z_1^*$. Extension to any $\bar{z}_2 \ne
z_1^*$ is possible since $\langle \sigma(z_1) \sigma(\bar{z}_2) \rangle_h$
\textit{is a function of} $z_1 - \bar{z}_2$. As the system is not
translationally invariant in space due to the
boundary, this is a highly non-trivial result. However, we have proved
this surprising property to all orders in $h$, justifying the
analytic continuation of Eq.~\eqref{M} to obtain $\langle \psi_{\sigma \alpha}(z_1) \psi_{\sigma' \beta}^\dagger(\bar{z}_2) \rangle
\propto \frac{\delta_{\sigma \sigma'}  \delta_{\alpha \beta}}{(z_1-\bar{z}_2)^{\frac{7}{8}}} m \left(\frac{-i z_1 +i \bar{z}_2}{2} \right)$. The special case of Eq.~\eqref{exactresult} with
finite $\lambda_1$ only follows by normalization and Fourier
transformation of this result.

\textit{Generalization to arbitrary relevant perturbation.--}
Generalizing the analysis of Ref.~\cite{SelePRL} to the 2IK  model
with an arbitrary combination of perturbations $\{\lambda_j\}$, it can
be shown that the 2IK FP Hamiltonian becomes SO$(8)$ symmetric:
$H_0=\frac{i}{2} \sum_{j=1}^8 \int_{-\infty}^\infty dx
\chi_{j}(x) \partial_x \chi_j(x)$, where $\{\chi_j \}$ are the 8
MFs. Switching on relevant perturbations at the critical point is
equivalent to adding $\delta H_{QCP}= i \sum_{j=1}^8
\lambda_j \chi_j(0) a$, which chooses  one direction in the
8-dimensional space. Defining a new
basis in which only $\chi(x) = \sum_j \lambda_j \chi_j(x) / \lambda$
couples to the local operator $a$, yields a Hamiltonian of the same
form as Eq.~\eqref{fieldtheorylimit}, with $h \to \lambda$ and $\epsilon
\to \chi$.
The full Green function for the 2IK model with generic relevant
perturbation can now be related to the result derived above for finite
$\lambda_1$ only. The required rotation in SO$(8)$ space is defined by
the unitary transformation $\chi \to U \chi U^\dagger = \chi_1$ with
$U = \exp \left(\gamma \int_{-\infty}^\infty dx \chi_1(x)
  \chi_\perp(x)  \right)$, where $\gamma=\arcsin
\frac{\lambda_\perp}{\lambda}$, $\lambda_\perp =\sqrt{\lambda^2 -
  \lambda^2_1}$, and $\chi_\perp = \sum_{j \ne 1} \lambda_j \chi_j /
\lambda_\perp$. The key point is that the
same rotation defined for the MFs can be used for the original
fermions since linear relations exist between their quadratic forms.
The general result for the 2IK model is Eq.~\eqref{exactresult}.

\emph{Extension to 2CK model.--} The same SO$(8)$ representation of the
critical point is obtained in both 2IK and 2CK models~\cite{maldacena}. Indeed, the CFT relevant
perturbations can be matched to MFs in the 2CK model as they were for
the 2IK model. Thus the above calculation can be generalized to
the 2CK model, with the key results given in
Eqs.~\eqref{tstar}--\eqref{Smatrix}.

\emph{Conclusion.--} We have derived a single exact crossover Green
function to describe the low-frequency crossover of the 2CK and 2IK
models. All relevant perturbations are related by an emergent SO$(8)$
symmetry, and should be regarded on an equal footing for $T^{*}\ll
T_K$, since marginal and irrelevant corrections to the critical point can then be
safely neglected~\cite{malecki}. The derivation depends on a non-trivial analogy between
the renormalization group flow in the 2IK model and in the BIM~\cite{CFT2IKM}, a
proof of which is the essentially perfect agreement between the exact
result and the full numerical solution obtained by NRG.

We acknowledge useful discussions with I. Affleck, P. Calabrese, D. E. Logan, A. Rosch, B. Rosenow, D. Schuricht and A. Stern. This work was supported by the
A.V. Humboldt Foundation (ES) and the DFG through SFB 608 and FOR 960
(AKM and LF).

\end{document}